# Embedded two-photon polymerized features with near-perfect lattice registration within self-organized photonic crystals and their optical properties

Erik C. Nelson, Florencio García-Santamaría, and Paul V. Braun*
*Department of Materials Science and Engineering, Frederick Seitz Materials Research Laboratory and Beckman Institute,*
*University of Illinois at Urbana-Champaign, Urbana, Illinois 61801, USA*

Two-photon polymerization has been demonstrated as an effective technique to define embedded defects in three-dimensional photonic crystals. In this work we demonstrate the ability to precisely position embedded defects with respect to the lattice of three-dimensional photonic crystals by imaging the structure concurrently with two-photon writing. Defects are written with near-perfect lattice registration and at specifically defined depths within the crystal. The importance of precise defect position is demonstrated by investigating the optical properties of embedded planar cavities written in a photonic crystal. The experimental data is compared to spectra calculated using the Scalar Wave Approximation (SWA) which further demonstrates the importance of defect placement.

## I. Introduction

Photonic crystals (PC's) are structures having a periodic modulation of dielectric constant on length scales comparable to the wavelength of light they are designed to manipulate. For certain three-dimensional (3D) structures of sufficiently high dielectric contrast, a photonic band gap (PBG) may result, preventing propagation of certain light frequencies within the crystal.[1-3] Many applications have been proposed for photonic band gap materials including optical circuits,[4-6] low-loss waveguides,[2,7] low-threshold lasers,[3] and sensors.[8] To realize many of these examples, simple periodic structures are not sufficient, and engineered defects (e.g. optical cavities) must be controllably positioned within the photonic crystal as well.[9]

While self-assembled colloidal crystals provide a simple approach to rapidly fabricating photonic crystals, the introduction of embedded defects requires additional processing, in contrast to direct fabrication approaches such as lithography or direct-laser writing where the defects are formed at the same time as the photonic crystal.[9-12] We have shown previously that two-photon polymerization (TPP) allows for the controlled addition of high resolution defects into a 3D self assembled colloidal crystal.[13-15] However many optically functional structures require not only high resolution features, but also exact placement of these features with respect to the photonic crystal lattice.[12,16-18] For example, the frequency of certain defect modes within three-dimensional PC's have been shown to be a function of defect position within the photonic crystal lattice.[17] Defect position also affects the coupling efficiency between, for example, a cavity and a waveguide or second cavity.[18] In addition to the position of the defect relative to the lattice, the proximity to defects inherent to the self-assembly process must also be considered since such intrinsic defects may degrade the optical properties of the system. While the intrinsic defect density of colloidal crystals has been substantially reduced in the past few years[19] there is a still a non-zero defect density which must be considered. Our previous demonstrations of TPP writing[13-15] relied on reflectance imaging to register the feature with the photonic crystal lattice. While this enabled general placement of features, reflectance imaging did not provide the resolution necessary to either avoid defects in the colloidal crystal (e.g. vacancies and stacking faults) or to register the feature with high resolution with respect to the crystal lattice. In some cases intrinsic defects could possibly be used to an advantage, for example vacancies may operate as optical cavities, and placement of features near such defects was not previously possible.

Here we demonstrate the ability to write embedded features with excellent lattice registration and placement with respect to intrinsic crystalline defects. Measurement and modeling of the variation of optical response for a planar defect as a function of its position within the crystal makes clear the need for this control. Concurrent high resolution imaging and feature writing was achieved by using a BODIPY fluorescent dye, for imaging that does not interact with the TPP system. BODIPY does not quench in the presence of our monomer, photons at its excitation wavelength are not absorbed by the photoinitiator, and no energy transfer from the dye to the multiphoton photoinitiator or monomer is observed. Consequently, it enables high resolution imaging of the photonic crystal lattice during TPP, facilitating exact placement features in well-ordered regions of a colloidal crystal.

## II. Reflectance vs. fluorescence imaging

Reflectance confocal imaging relies on reflections from interfaces to construct an image. By necessity, for TPP, the monomer is nearly index matched to the colloids to minimize light scattering during writing, therefore reflections from the monomer-colloid interface are low reducing the resolution of the reflectance image. At the interface of a strongly reflecting substrate such as silicon, a fluorescence image provides only slightly more information than a reflectance image **(Fig. 1a,b)**. Both images show a line defect in the lower right corner of the image. In the reflectance image, the colloids are bright, while in the fluorescence image the colloids are dark as the dye is contained in the monomer.

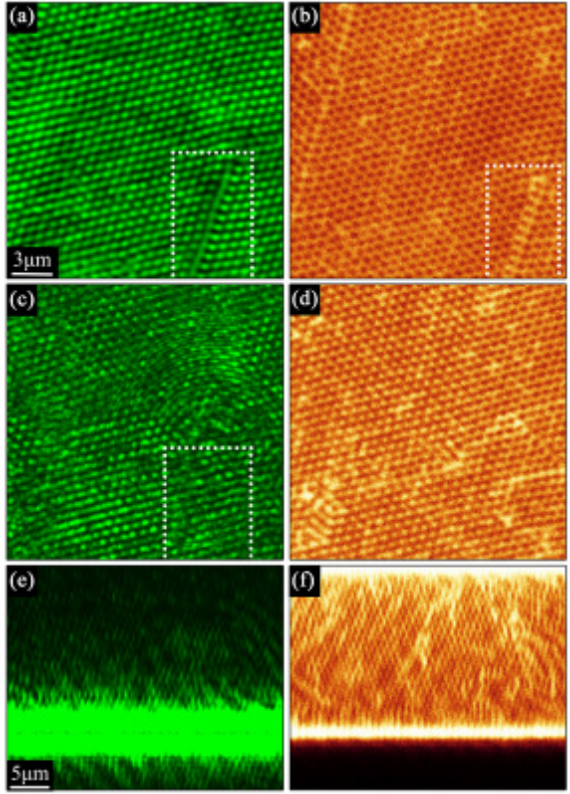

**Figure 1.** Laser scanning confocal microscope images. The first layer of colloids on the substrate surface: (a) reflectance and (b) fluorescence imaging; dotted boxes highlight a line defect. Near the midplane of the colloidal crystal, parallel to the substrate: (c) reflectance (dotted box highlights image artifact of line defect) and (d) fluorescence imaging. Cross-section: (e) reflectance and (f) fluorescence imaging. (a-d) are the same scale and (e, f) are the same scale.

Deeper in the crystal, reflectance imaging is not effective; the images are artifacts of strong reflections from the colloidal crystal-substrate interface (Fig. 1c). The line defect from the first layer is still apparent while the fluorescence image of the same plane of the crystal (Fig. 1d) shows it is an artifact. In addition, the intrinsic defects seen in Figure 1d are not apparent in the reflectance image, Reflectance artifacts are clearly seen in reflectance images perpendicular to the substrate (Fig. 1e) where colloids are observed on both sides of the substrate-colloidal crystal interface, even though they only exist on one side of the substrate. Fluorescence imaging enables complete 3D imaging of the crystal, including identification of intrinsic defects (Fig. 1f). This complete 3D structural information is necessary for positioning of optically functional features.

### III. Two-photon polymerization of registered features

TPP features can easily be written through the approximately 17 μm thickness of the colloidal crystal **(Fig. 2a-c)** using fluorescence imaging (Fig. 2c) for alignment. SEM images (Figs. 2a,b) confirm the high degree of registration seen in the fluorescence image. Equally important, registration can be maintained in multiple crystallographic directions in this case through a 60 degree bend.

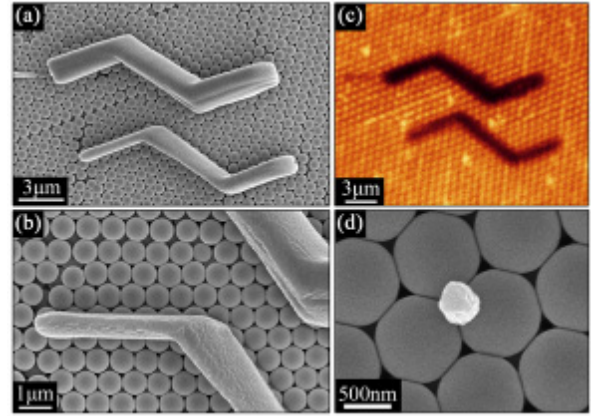

**Figure 2.** a) SEM image of two TPP features drawn through the top of the crystal showing the high degree of registration with the colloidal crystal lattice. b) SEM image of the narrower TPP feature highlighting the registration along two crystallographic directions. c) Laser scanning fluorescence confocal microscope image parallel to the substrate of TPP features. The image plane is within the crystal. Defects can be seen near the TPP feature as bright dots or lines. d) SEM image of a TPP point feature placed at the interstitial space between three colloids.

The SEM images also confirm the ability of fluorescence imaging to detect crystalline defects. The fluorescence image obtained near the top of the crystal (Fig. 2c) shows a bright line defect running between the first positively sloped section of the two features as well as several point defects. The line defect, which propagates up to the top surface of the crystal, is also visible in the SEM image (Fig. 2a). In addition to straight line features, point features smaller than 500nm in diameter can also be precisely positioned. Figure 2d shows a point defect placed in the interstitial site of the colloidal lattice. A number of designs for optically active structures in PCs rely on precise placement of point defects[16,18,20-22] and, due to their small size relative to the PC structure, precise control over their position is very important.

While aligning features written through the crystal is an effective demonstration of our technique, it is important for proposed optical devices that registration can be achieved for *embedded* features. Here we demonstrate that fluorescence imaging enables registration of embedded TPP features with the colloidal crystal lattice and the avoidance of intrinsic defects **(Fig. 3)**. Figure 3a presents an embedded polymer feature written with lattice registration near two point defects, which can be observed as bright spots to the right and left of the middle segment of the feature. While intrinsic defects may have a negative effect on optical properties, this feature was intentionally written between the intrinsic defects to demonstrate the control afforded by our technique. Figure 3b shows that the feature was written within the bulk of the colloidal crystal, not at an interface where reflection imaging might still be possible.

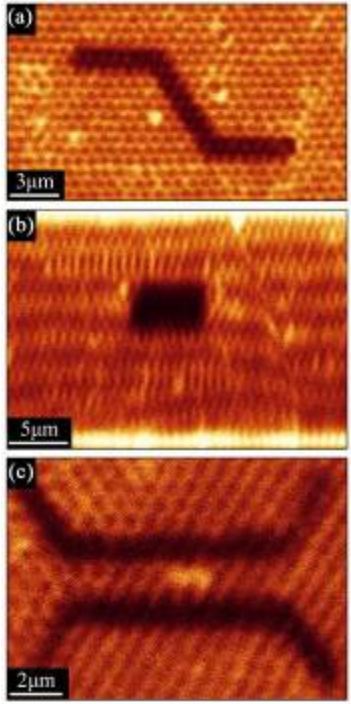

**Figure 3.** Laser scanning fluorescent confocal microscope images of embedded polymer features. a) Parallel to the substrate. Two point defects in the colloidal crystal are seen as regions of high fluorescent intensity adjacent to the feature. b) Perpendicular to the substrate. The feature is embedded near the midplane of the colloidal crystal. c) Parallel to the substrate. Two waveguide features written adjacent to a double point defect.

The ability to register embedded features with the lattice and position them accurately with respect to crystalline defects makes possible the concept of using the intrinsic defects as optical elements. Figure 3d presents an example of two waveguide structures positioned on either side of a double point defect. A number of proposed PC devices rely on such waveguide-cavity structures,[21-24] including proposed designs for channel drop filters.[25,26] The proposed structures require the waveguides to be positioned within a few lattice constants of the resonant cavity. Figure 3d demonstrates a simple fabrication route for such a device, using the intrinsic point defects of the colloidal crystal as the source of the resonant cavity. The size of such a cavity can be tuned by various processing techniques we have described elsewhere[14,27] to tune the optical properties. The waveguides are written within a lattice constant of either side of the cavity, demonstrating the control required according to theoretical studies.[25,26]

### IV. Effect of feature placement on optical properties

While the crystal is effectively infinite in the x-y plane, this is not the case in the z-direction where it is expected that the macroscopic position of the defect, not just its position in the unit cell, will affect the optical properties. This is especially important for cases where light is coupling to an embedded cavity. In this section we show that the precise placement of planar cavities with respect to the crystal thickness has a strong effect on both the spectral width of the defect mode and its intensity.

Fluorescence imaging and our TPP fabrication process allow the precise positioning of defects in the crystal thickness; the optical properties of these defects were subsequently interrogated. Planar cavities were fabricated **(Fig. 4b, c)** positioned at 0.25, 0.50, and 0.75 of the total crystal thickness (measured from the substrate, Fig. 4a). The precise alignment of the planar defects within the PC can be seen clearly in Figures 4d-f. To demonstrate the importance of feature placement in the direction of crystal thickness on the optics of defects, the reflectance was measured for the 0.25, 0.50 and 0.75 planar defects and compared to theoretical reflectance data **(Fig. 5)**. Simulation data are Scalar Wave Approximation (SWA) plots with the cavity modeled as an opal infiltrated with polymer. The theory matches the data quite well for each case. The defect mode moves from higher to lower energy as the defect is positioned further from the substrate, both in simulation and experiment. The highlighted gray region is the range of frequencies the defect modes occupy for the various defect placements. For this cavity thickness the defect modes appear at the high energy edge of the gap (the position of the defect within the gray box is a guide to the eye to see where in the bare opal reflectance peak the defect mode appears).

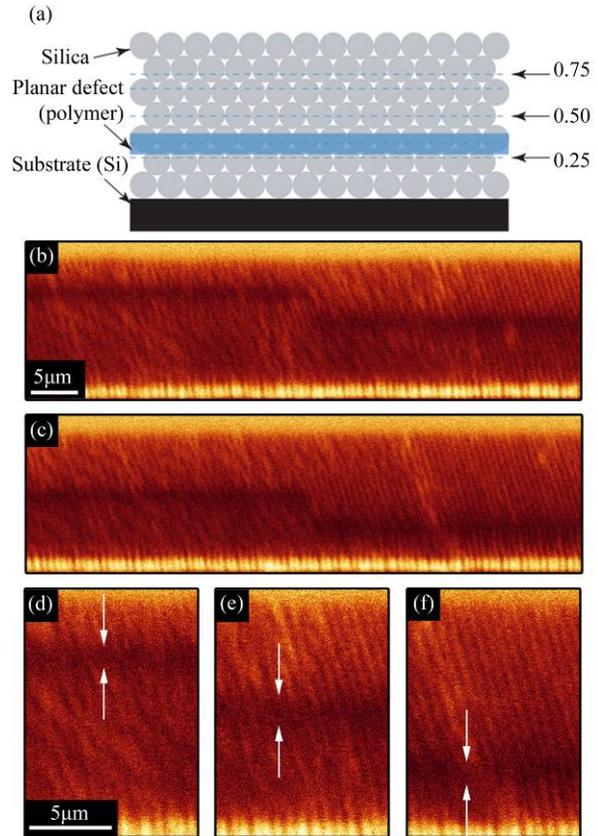

**Figure 4.** a) Schematic of planar defects written at various locations within a colloidal PC. b) Defects written at 0.75 and 0.50 according to the schematic in (a). c) Defects written at 0.50 and 0.25 according to the schematic in (a). Magnified images of defects written at 0.75 (d), 0.50 (e) and 0.25 (f) – the white arrows are guides for the eye.

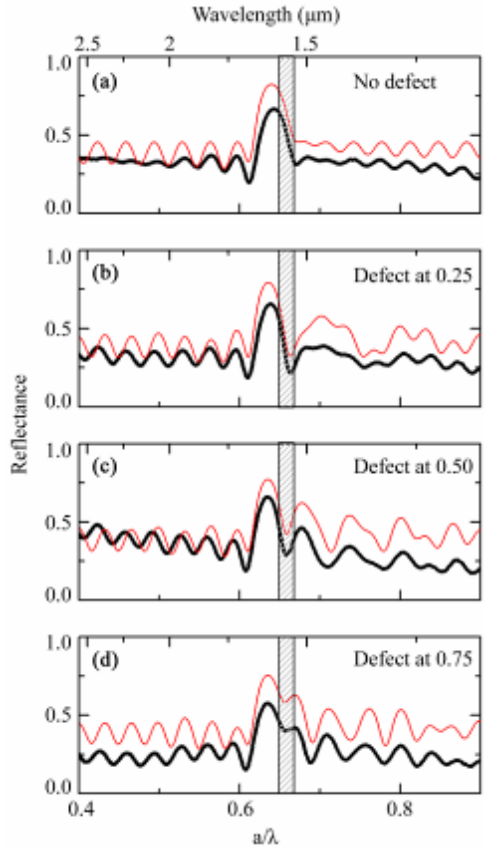

**Figure 5.** Experimental (thick line) and simulated (thin line) reflectance data for a bare opal (a), opal with a defect at 0.25 of the crystal thickness (b), defect at 0.50 of the crystal thickness (c) and defect at 0.75 of the crystal thickness. The gray box highlights the range of frequencies where the various defect modes appear.

Intuitively, one may expect symmetry across the plane of the center of the crystal yielding identical reflectance for cavities at, for example, 0.75 and 0.25. This is the case for a crystal surrounded by air on both sides, as seen in Figure 6a. The contour plots in **Figure 6** are defect position versus frequency with the colors representing reflectance. When there is a substrate in the system symmetry is broken and now cavities placed at 0.75 and 0.25 demonstrate different behavior (Fig. 6b-d). The refractive index of the substrate dictates the degree of asymmetry of the system; a crystal on glass (Fig. 6b) is far more similar to the symmetric air/crystal/air case than a crystal on silicon (Fig. 6c). Photonic crystals designed for in the infrared are typically grown on silicon, as would crystals incorporated into an integrated on-chip optical system. Therefore the defect position should be decided on and fabricated carefully.

Cavity placement appears to have a strong effect on the "strength" of the modes (defined for this case as the decreased reflectance compared to the bare opal case for the defect mode frequencies). The cavity at 0.25 demonstrates a broad mode and, though the strength of the mode seems large, the peak is blue shifted and is further down the reflectance peak making the actual change in reflectance small compared to the bare opal case (Fig. 5b). This is expected due to the low crystal thickness on one side of the cavity making it difficult to build up a resonance. In addition the thin crystal side of the cavity terminates at the substrate which causes a further change in the mode shape. This is seen most clearly in Fig 6c which also shows that as the cavity approaches the substrate the resonance disappears entirely and there is no defect mode.

Similar behavior is seen for the cavity at 0.75 (Fig. 5d), though in this case the mode is much less broad. The reflectance is at a larger magnitude since this defect mode is red shifted and lies farther up the reflectance peak (though the strength of the mode is similar to the 0.25 cavity). The cavity at 0.5 shows the greatest strength of the mode since it has nearly equal thickness of crystal on both sides of the cavity and has a greater ability to build up a cavity resonance. The varying behavior with cavity placement is most clearly visualized in Figure 6c. It is clear that cavities placed near the center of the crystal exhibit the greatest mode strength and that the width of the defect modes varies rapidly as the cavity is moved up or down from the center of the crystal. This makes defect placement important if a certain mode bandwidth is required, which is often the case. The case of a 40 layer crystal was also simulated to show the effects of stronger confinement within a cavity (Figure. 6d). The width of the modes is greatly decreased in this case though the asymmetric behavior of defect placement is still present. Defects placed in the center of the crystal continue to show the greatest mode strength. However the broadening for cavities closer to the substrate is now greatly reduced and the mode strength somewhat improved. In contrast the mode strength for defects near the top of the crystal becomes very poor. With increased crystal thickness, though, there is increased tolerance for defect placement to achieve the same or very similar mode strength and width. In high quality thin film opals typical thicknesses are around 25 layers or less so defect placement is particularly important.

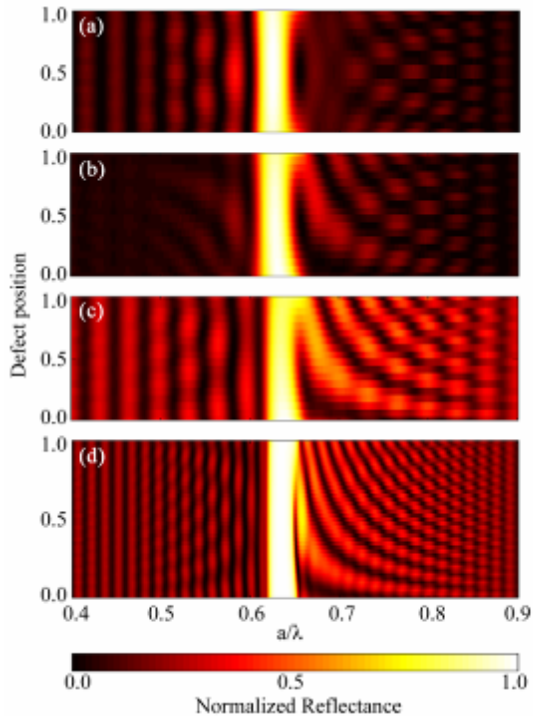

**Figure 6.** Simulated reflected intensity as a function of defect position and normalized frequency for 17 layer opals with: (a) air substrate; (b) glass substrate; (c) Si substrate. A 40 layer opal with Si substrate is shown in (d). All reflectance values are normalized to the maximum and minimum for each simulation for the color bar.

## V. Conclusions

The advantages of fluorescence imaging in conjunction with TPP are clearly apparent. Imaging through the thickness parallel to the substrate, imaging the entire cross-section perpendicular to the substrate as well as direct observation of intrinsic crystalline defects have been successfully demonstrated. Fluorescence imaging provides great flexibility in positioning embedded defects by enabling precise control over location with respect to intrinsic crystal defects as well as feature registration with the PC lattice. The optics of embedded features can be greatly affected by not only their in-plane alignment but also their position relative to the surface of the PC. The precise positioning of defects in the crystal thickness and the corresponding effects on their optical properties have been demonstrated. It is seen that the ability of light to couple to embedded defects and build resonance as well as the spectral width of the defect modes are highly dependant on the location of the defect within the crystal thickness, making precise control over defect placement essential. The techniques demonstrated here may make possible numerous photonic devices which require specific feature placement within the photonic structure.

## VI. Acknowledgements

This material is based upon work supported by the U. S. Army Research Laboratory and the U. S. Army Research Office grant DAAD19-03-1-0227. This work was carried out in part in the Beckman Institute Microscopy Suite, UIUC and the Center for Microanalysis of Materials, UIUC, which is partially supported by the U.S. Department of Energy under grants DE-FG02-07ER46453 and DE-FG02-07ER46471. We gratefully acknowledge Dr. L.-S. Tan (U.S. Air Force Research Laboratory) for providing the two-photon sensitive dyes and Dr. S. A. Rinne (UIUC) for experimental assistance.
*Electronic address: pbraun@uiuc.edu

## VII. Experimental Methods

Colloidal crystals were prepared from 730nm and 920nm diameter silica particles using a vertical evaporation technique similar to that described elsewhere.[19] Opals were filled with 6nm of aluminum oxide grown conformally using Atomic Layer Deposition (Cambridge Nanotech Savannah). Colloidal crystals were characterized by FTIR spectroscopy using a Bruker Hyperion microscope coupled into a Bruker Vertex 70 FTIR spectrometer equipped with a 4x, 0.1 NA objective and a collection area of 187.5um achieved using a spatial aperture. Crystal quality was evaluated by measuring the lowe energy reflectance peak caused by reflections from opal layers in the (111) direction in several locations on the crystal. The crystals used in this study had reflectance values ranging from 63-81% for 15 to 22 layers.

The TPP solution was composed of trimethylolpropane triacrylate (TMPTA) monomer (Sigma Aldrich) with inhibitor; 0.1 wt% AF-350 photoinitiator (tris[4-(7-benzothiazol-2-yl-9,9-diethylfluoren-2-yl)phenyl]amine) donated by the Air Force Research Laboratory[28] and [10 μM] Invitrogen BODIPY 630/650-X, SE (6-(((4,4-difluoro-5-(2-thienyl)-4-bora-3a,4a-diaza-*s*-indacene-3-yl)styryloxy)acetyl)aminohexanoic acid, succinimidyl ester). The BODIPY dye has an absorption maximum of 625nm and an emission maximum of 640nm; neither wavelength causes single-photon excitation of the photoinitiator.

Confocal imaging and TPP were performed on a laser scanning confocal microscope (Leica DMIRBE with an SP2 scanhead). Reflectance and fluorescence imaging was performed with a 633nm HeNe laser. Spatially defined TPP was achieved through our published procedure.[13,14] Point defects were written using software defined bleach points. Samples were mounted with the colloidal crystal facing a coverslip; TPP is performed through the coverslip. Imaging and ROI alignment were performed using a beam expander to fill the back aperture of the objective to improve image quality in the microscope. TPP was performed with the beam expander removed; rotational alignment was unaffected by the change because the scan field does not rotate. The absolute position of the ROI in the x-y plane (parallel to the substrate) varies with the presence beam expander and is compensated for prior to writing the feature. The position of the ROI and the resultant polymer feature are systematically not coincident in the direction of the crystal thickness; the ROI is positioned offset from the desired feature location to compensate. After TPP samples were rinsed in ethanol to remove excess monomer and dried in air. Scanning electron microscope (SEM) samples were gold-palladium coated prior to imaging; the micrographs were taken using a Hitachi S-4700 SEM.

The parameters for the SWA simulations are: $n_{silica}$=1.44, $n_{polymer}$=1.6, $n_{air}$=1; 12 layers above and 5 below (0.25 cavity), 8 layers above and 9 below (0.5 cavity) and 4 layers above 13 below (0.75 cavity). Layers above and below do not equal exact crystal thickness fractions because the SWA code requires the crystal thicknesses be integer numbers of layers; the thickness fractions are within a few percent and the fits are good. In order to fit the data the SWA simulations required 17 total layers of crystal (sum of layers above and below the cavity) for the cases of the defect and 18 layers for the bare opal (without a cavity). This can be understood by the fact that the thickness of the total system for the bare opal is simply the layers of the crystal whereas the cases with the cavity also include the thickness of the defect (1.12a in this case, or around one layer) so the total thickness of each system is approximately the same. The spheres are interpenetrated to model the experimental system where a thin ALD layers is grown on the spheres to yield interpenetration.